\documentclass[aps,pre,amsmath,amssymb,twocolumn,groupedaddress,showpacs,eqsecnum]{revtex4}
\usepackage{graphicx}
\usepackage[utf8]{inputenc}

\usepackage{epsf}

\newcommand{\be}{\begin{equation}}
\newcommand{\ee}{\end{equation}}
\newcommand{\ba}{\begin{eqnarray}}
\newcommand{\ea}{\end{eqnarray}}
\newcommand{\baa}{\begin{eqnarray}}
\newcommand{\eaa}{\end{eqnarray}}
\newcommand{\ed}{\end{document}}

\renewcommand{\baselinestretch}{1.2}
\date{\today}

\begin{document}
\title{Scheme for accelerating quantum tunneling dynamics}
\author{ Anvar Khujakulov$^{(1)}$ and Katsuhiro Nakamura$^{(2,3)}$}
\affiliation{$^{(1)}$Department of Physics, Heinrich-Heine-University D\"usseldorf, Universit\"at strasse 1, 40225 D\"usseldorf, Germany\\
$^{(2)}$Faculty of Physics, National University of Uzbekistan, Vuzgorodok, Tashkent 100174, Uzbekistan\\
$^{(3)}$Department of Applied Physics, Osaka City University, Sumiyoshi-ku, Osaka 558-8585, Japan\\
}
\begin{abstract}
We propose a scheme of the exact fast-forwarding of
standard quantum dynamics for a charged particle. The present idea allows the acceleration of both the amplitude and phase of the wave function throughout the fast-forward time range and is distinct from that of Masuda-Nakamura (e.g., Proc. R. Soc. A {\bf 466}, 1135
(2010)) which enabled acceleration of only the amplitude of the wave function on the way.
We shall apply the proposed method to the quantum
tunneling phenomena and obtain the electro-magnetic field to ensure the rapid penetration of wave functions through
a tunneling barrier. Typical examples described here are: 1) an exponential wave packet passing
through the delta-function barrier;  2)  the
opened Moshinsky shutter with a delta-function barrier just behind
the shutter.
We elucidate the tunneling current in the vicinity of the barrier and find the remarkable enhancement of the tunneling rate (: tunneling power) due to the fast-forwarding. In the case of a very high barrier, in particular, we present the asymptotic analysis and exhibit a suitable driving force to recover a recognizable tunneling current. The analysis is also carried out on the exact acceleration of macroscopic quantum
tunneling with use of the nonlinear Schr\"odinger equation which
accommodates a tunneling barrier .
\end{abstract}
\maketitle

\section{Introduction}\label{sec-introduction}
One of the most fascinating phenomena in quantum mechanics is a
quantum tunneling through a barrier. The tunneling shows up in
Zener tunneling in biased semiconductors, quantum devices like diodes, scanning tunneling microscopy, $\alpha$ decay in heavy nuclei, etc.  In general, however, the tunneling rate (: tunneling power) is very small or the tunneling time is very long. Even in the case of resonant tunneling of electrons through hetero-structures \cite{esaki}, there is room for research on the tunneling time \cite{stein}. Therefore it is desirable to invent a protocol of accelerating
the tunneling.

Masuda and Nakamura \cite{mas1,mas2,mas3} investigated a way to
accelerate quantum dynamics with use of a characteristic driving
potential determined by the additional phase of a wave function.
One can accelerate a given quantum dynamics to obtain a
target state in any desired short time. This kind of acceleration
is called the fast forward \cite{term} of quantum dynamics, which
constitutes one of the promising ways of the shortcut to adiabaticity
\cite{dr1,dr2,mb,lr,mg1,mg2,mnc}. The relationship between the fast forward and the
shortcut to adiabaticity is nowadays clear\cite{mg-nice,tk-nice}.

The idea of the
tunneling seems to be incompatible with that of fast forward. But
here one can combine these two ideas, that is, one can conceive a theory to
accelerate the tunneling dynamics through the high barrier and complete,
in any desired short time, the tunneling phenomena which
originally needs a long tunneling time.

Before embarking upon the main part of the text, we briefly summarize the
previous theory of the fast-forward of quantum dynamics by Masuda
and Nakamura \cite{mas1}. The Schr\"{o}dinger equation in standard time with a nonlinearity constant $c_0$ (appearing in macroscopic quantum dynamics) is
represented as
\begin{eqnarray}\label{MN-standeq}
\imath\hbar\frac{\partial{\psi_{0}}}{\partial{t}}=-\frac{\hbar^{2}}{2m}\nabla^{2}\psi_{0}
+V(\textbf{x},t)\psi_{0} -c_0|\psi_0|^2\psi_0.
\end{eqnarray}
$\psi_{0} \equiv \psi_0(\textbf{x},t)$ is a known function of
space $\textbf{x}$ and time $t$ under a given potential $V(\textbf{x},t)$ and is called a standard state. For any
long time $T$ called as a standard final time, we choose
$\psi_{0}(t=T)$ as a target state that we are going to generate.

Let ${\tilde{\psi}_0}(\textbf{x},t)$ be a fast-forwarded state
of $\psi_{0}(\textbf{x},t)$ as defined by
\begin{eqnarray}
\tilde{\psi}_0(\textbf{x},t)\equiv \psi_{0}(\textbf{x},\Lambda(t))
\end{eqnarray}
with
\begin{eqnarray}\label{ff-t}
\Lambda(t)=\int_0^t\mathrm{\alpha(t')}\,\mathrm{d}t',
\end{eqnarray}
where $t$ is a new time variable distinct from the standard one.
$\alpha (t)$ is a magnification time-scale factor defined by
\begin{eqnarray}\label{alpha}
\alpha(0) &=&1,\nonumber\\
\alpha(t) &>&1    \qquad (0<t<T_{FF}),\nonumber\\
\alpha(t)&=&1    \qquad (t \ge T_{FF}).
\end{eqnarray}
We consider the fast-forward dynamics with a new time variable which reproduces the target state $\psi_0(T)$
in a shorter final time $T_{FF}(<T)$
defined by
\begin{eqnarray}\label{T-Tff}
T=\int_0^{T_{FF}}\alpha (t) \mathrm{d}t.
\end{eqnarray}

Since the generation of $\tilde{\psi}_{0}$ requires an anomalous mass
reduction, $\tilde{\psi}_{0}$ as it stands cannot be a candidate for the fast-forward state\cite{mas1} .
But one can obtain the target state by considering a fast-forwarded
state $\psi_{FF}=\psi_{FF}(\textbf{x},t)$ which differs from
$\tilde{\psi}_0$ by an extra phase as
\begin{eqnarray}\label{add-f}
\psi_{FF}(t)=e^{{\rm i}f}\tilde{\psi}_0(t)=e^{{\rm i}f}\psi_0(\Lambda(t))
\end{eqnarray}
where $f\equiv f(\textbf{x},t)$ is a real
function of  $\textbf{x}$ and  $t$ and is called the additional phase.
With use of a new time variable $t$ that appeared in Eq.(\ref{ff-t}),
the Schr\"{o}dinger equation for fast-forward state $\psi_{FF}$ is
supposed to be given as
\begin{eqnarray}\label{MN-ffeq}
{\rm i}\hbar\frac{\partial{\psi_{FF}}}{\partial{t}}=-\frac{\hbar^{2}}{2m}\nabla^{2}\psi_{FF}
+V_{FF}\psi_{FF} - c_0|\psi_{FF}|^2\psi_{FF}\nonumber\\
\end{eqnarray}
with the driving scalar potential $V_{FF}=V_{FF}(\textbf{x},t)$.
If we shall choose $\alpha(t)$ as in Eq.(\ref{alpha}),
the additional phase can vanish at the final time of the
fast-forward $(T_{FF})$ and we can obtain the exact target state
\begin{eqnarray}\label{TFF-T}
\psi_{FF}(T_{FF})=\psi_{0}(T).
\end{eqnarray}

The explicit expression for $\alpha(t)$ in the fast-forward range ($0 \le t \le T_{FF}$) is  proposed by Masuda and Nakamura\cite{mas1,mas2,mas3} as:
\begin{eqnarray}\label{nonunif-scale}
\alpha(t)=(\bar{\alpha}-1)\cos\left(\frac{2\pi}{T/\bar{\alpha}}t+\pi\right)+\bar{\alpha},
\end{eqnarray}
where $\bar{\alpha}$ is the mean value of $\alpha(t)$ and is given by $\bar{\alpha}=T/T_{FF}$.
Besides the time-dependent scaling factor in Eq.(\ref{nonunif-scale}) in the fast-forward range, we can also have recourse to the uniform scaling factor :
\begin{eqnarray}\label{unif-scale}
\alpha(t)=\bar{\alpha} \qquad  ( 0\le t \le T_{FF}),
\end{eqnarray}
which will be useful in the quantitative analysis of fast forward.
Substituting Eqs.(\ref{MN-standeq}),(\ref{ff-t}) and (\ref{add-f}) into Eq.(\ref{MN-ffeq}) and taking its real
and imagery parts, we obtain a pair of equations for $f$ and $V_{FF}$, which are solvable.

While the above idea guarantees the exact target state at $t=T_{FF}$, in the intermediate time range $0\le t \le T_{FF}$ it accelerates only the amplitude of the wave function and fails to accelerate its phase because of the non-vanishing additional phase $f$ in Eq.(\ref{add-f}) on the way.
If one wish to accelerate the time-dependent current, one must innovate the theory to recover the phase exactly in the intermediate time range until $t=T_{FF}$, which will be done below.
\renewcommand{\labelenumi}{\theenumi.}

Our theory will hold to both quantum  dynamics described by the Schr\"odinger equation and macroscopic quantum dynamics described by the nonlinear Schr\"odinger equation. Section \ref{new appro} is concerned with a new framework of the exact fast forward of both quantum and macroscopic quantum dynamics. Sections \ref{FF-TUNNEL-WP}  and \ref{Moshinsky} are devoted to application to quantum tunneling ($c_0=0$)
and Section \ref{macro} treats the macroscopic quantum tunneling which includes the nonlinearity (
$c_0 \ne 0$).
 \section{New approach to fast-forward theory}\label{new appro}

The Schr\"{o}dinger equation for the wave function $\psi_{0} \equiv \psi_0(\textbf{x},t)$ of a charged particle in the presence of the scalar potential  $V$ is the same as in Eq.(\ref{MN-standeq}).
For any long time $T$ called as a standard final time, we choose
$\psi_{0}(\textbf{x},T)$ as a target state that we are going to generate.
In contrast to the previous works \cite{mas1,mas2,mas3}, the fast-forward wave function here does not include the additional phase factor throughout the fast forwarding time range until $T_{FF}$ in Eq.(\ref{T-Tff})
and is given by
\begin{eqnarray}\label{ffwf}
\psi_{FF}(\textbf{x},t)=\psi_{0}(\textbf{x},\Lambda(t))=\tilde{\psi}_{0}(\textbf{x},t).
\end{eqnarray}
Here $\Lambda(t)$ is the same as in Eq.(\ref{ff-t}).
We shall try to realize $\psi_{FF}$ by applying the electro-magnetic field $\textbf{E}_{FF}$
and $\textbf{B}_{FF}$.

Let's assume $\psi_{FF}$ to be the solution of Schr\"{o}dinger
equation for a charged particle in the presence of  additional vector $\textbf{A}_{FF}(\textbf{x},t)$ and scalar
$V_{FF}(\textbf{x},t)$ potentials, as given by
\begin{eqnarray}\label{ffeq}
\imath\hbar\frac{\partial{\psi_{FF}}}{\partial{t}}&=&
\left(\frac{1}{2m}\left(\frac{\hbar}{i}\nabla-\frac{q}{c}\textbf{A}_{FF}\right)^2+ q V_{FF}
+V\right)\psi_{FF}\nonumber\\
&-& c_0|\psi_{FF}|^2\psi_{FF}\nonumber\\
&=&-\frac{\hbar^{2}}{2m}\nabla^{2}\psi_{FF}
+\frac{i\hbar q}{2m c}(\nabla\cdot\textbf{A}_{FF})\psi_{FF}\nonumber\\
&&+\frac{i\hbar q}{m c} \textbf{A}_{FF}\cdot\nabla\psi_{FF}
+\frac{q^2\textbf{A}_{FF}^{2}}{2m c^2}\psi_{FF}\nonumber\\
&&+(q V_{FF}+V)\psi_{FF} - c_0|\psi_{FF}|^2\psi_{FF}.
\end{eqnarray}
For simplicity, however, we shall hereafter employ the unit of velocity of light $c=1$ and the prescription of a positive unit charge $q=1$. Note: $V_{FF}$
in Eq.(\ref{ffeq}) is introduced independently from a given potential $V$, in contrast to the one in Eq.(\ref{MN-ffeq}) which included $V$. The driving electro-magnetic field is
related by,
\begin{eqnarray}\label{EB-AV}
\textbf{E}_{FF}=-\frac{\partial \textbf{A}_{FF}}{\partial t}-\nabla V_{FF},\nonumber
\\\textbf{B}_{FF}=\nabla\times\textbf{A}_{FF}.
\end{eqnarray}
Substituting Eqs (\ref{MN-standeq}), (\ref{ffwf}) into Eq.(\ref{ffeq}) and taking its
real and imaginary parts, we obtain a pair of equations
\begin{eqnarray}\label{aff}
\nabla\cdot\textbf{A}_{FF}&+&2\textrm{Re}\left[\frac{\nabla\tilde{\psi}_{0}}{\tilde{\psi_{0}}}\right]\textbf{A}_{FF}\nonumber\\
&+&\hbar(\alpha-1)
\textrm{Im}\left[\frac{\nabla^{2}\tilde{\psi}_{0}}{\tilde{\psi_{0}}}\right]=0
\end{eqnarray}
and
\begin{eqnarray}\label{vff}
V_{FF}&=&-(\alpha-1)\frac{\hbar^{2}}{2m}\textrm{Re}\left[\frac{\nabla^{2}\tilde{\psi}_{0}}{\tilde{\psi_{0}}}\right]\nonumber\\
&+&\frac{\hbar}{m}\textbf{A}_{FF}\textrm{Im}\left[\frac{\nabla\tilde{\psi}_{0}}{\tilde{\psi_{0}}}\right]
-\frac{1}{2m}\textbf{A}_{FF}^{2}+(\alpha-1)V\nonumber\\
&-&(\alpha-1)c_0|\tilde{\psi_0}|^2.
\end{eqnarray}
Now we write
$\tilde{\psi}_{0}$ as
\begin{eqnarray}
\tilde{\psi}_{0}=\rho e^{i\eta}
\end{eqnarray}
with use of the real amplitude $\rho$ and phase $\eta$
defined by
\begin{eqnarray}
\rho&\equiv&\rho(\textbf{x},\Lambda(t)),\nonumber\\
\eta&\equiv&\eta(\textbf{x},\Lambda(t)).
\end{eqnarray}

Then, using in Eqs.(\ref{aff}) the equalities $\textrm{Re}\left[\frac{\nabla\tilde{\psi}_{0}}{\tilde{\psi_{0}}}\right]=\frac{\nabla\rho}{\rho},\quad\textrm{Im}\left[\frac{\nabla\tilde{\psi}_{0}}{\tilde{\psi_{0}}}\right]=\nabla\eta,
\quad\textrm{Re}\left[\frac{\nabla^{2}\tilde{\psi}_{0}}{\tilde{\psi_{0}}}\right]=\frac{\nabla^{2}\rho}{\rho}-(\nabla\eta)^{2},\quad
\textrm{Im}\left[\frac{\nabla^{2}\tilde{\psi}_{0}}{\tilde{\psi_{0}}}\right]=2\frac{\nabla\rho}{\rho}\nabla\eta+\nabla^{2}\eta$, one finds that
\begin{eqnarray}\label{aff-sol}
\textbf{A}_{FF}=-\hbar(\alpha-1)\nabla\cdot\eta
\end{eqnarray}
satisfies Eq.(\ref{aff}).
Thanks to Eq.(\ref{MN-standeq}) with the variable $t$ being replaced by $\Lambda(t)$,
$\textrm{Re}\left[\frac{\nabla^{2}\tilde{\psi}_{0}}{\tilde{\psi_{0}}}\right]$ can be re-expressed  as
\begin{eqnarray}\label{2deri-rev}
\textrm{Re}\left[\frac{\nabla^{2}\tilde{\psi}_{0}}{\tilde{\psi_{0}}}\right]=\frac{2m}{\hbar}
\left(\frac{\partial \eta}{\partial \Lambda(t)}+\frac{1}{\hbar}V \right)-\frac{2mc_0}{\hbar^2}\rho^2.
\end{eqnarray}
Then $V_{FF}$ can be expressed only with use of $\eta$ as
\begin{eqnarray}\label{vff-sol}
V_{FF}&=&-(\alpha-1)\hbar\frac{\partial \eta}{\partial \Lambda(t)}\nonumber\\
&-&\frac{\hbar^{2}}{2m}(\alpha^2-1)(\nabla \eta)^2.
\end{eqnarray}
With use of the driving vector
$\textbf{A}_{FF}$ and scalar $V_{FF}$ potentials in
Eqs.(\ref{aff-sol}) and (\ref{vff-sol}), we can obtain the fast-forwarded state $\psi_{FF}$ in Eq.(\ref{ffwf})
which is now free from the additional phase factor $f$ in Eq.(\ref{add-f}) used in Masuda and
Nakamura's framework\cite{mas1,mas2,mas3}. Logically, $\textbf{A}_{FF}$ and  $V_{FF}$, which prove to be both independent of the amplitude $\rho$, serve to compensate the additional phase $f$ in Eq.(\ref{add-f}) in their framework.
The electro-magnetic field introduced in Refs. \cite{mas3,kiel} is designed to guarantee the equality
in Eq.(\ref{TFF-T}) at $t=T_{FF}$, and fails in removing the additional phase
in the fast-forward time range  $0<t<T_{FF}$.

Two points should be noted:
1) The above driving potentials do not explicitly depend on the nonlinearity coefficient $c_0$:
Eqs.(\ref{aff-sol}) and (\ref{vff-sol}) work for the nonlinear Schr\"odinger equation  as well.
2) The magnetic field ${\bf B}_{FF}$ is vanishing, because a combination of Eqs. (\ref{EB-AV})
and (\ref{aff-sol}) leads to
${\bf B}_{FF}={\bf \nabla} \times {\bf A}_{FF}=0$. Therefore only the electric field
$\textbf{E}_{FF}$ is required to accelerate  a given dynamics. With use of
Eqs. (\ref{EB-AV}), (\ref{aff-sol}) and (\ref{vff-sol}), $\textbf{E}_{FF}$ is given explicitly by
\begin{eqnarray}\label{ele-fld}
\textbf{E}_{FF}&=&\hbar \dot{\alpha}\nabla \eta+ \hbar \frac{\alpha^2-1}{\alpha}\partial_t\nabla\eta \nonumber\\
&+&\frac{\hbar^{2}}{2m}(\alpha^2-1)\nabla (\nabla \eta)^2.
\end{eqnarray}

A remarkable issue of the present scheme is the enhancement of the current density $\textbf{j}_{FF}$. Using a generalized momentum which includes a contribution from the vector potential in
Eq.(\ref{aff-sol}), we see:
\begin{eqnarray}\label{ff-curren}
\textbf{j}_{FF}(\textbf{x},t)&\equiv&
\textrm{Re}[\psi_{FF}^{*}(\textbf{x},t)\frac{1}{m}\left(\frac{\hbar}{i}\nabla - \textbf{A}_{FF}\right)
\psi_{FF}(\textbf{x},t)]\nonumber\\
&=&\frac{\hbar}{m}\alpha(t)\rho^2(\textbf{x},\Lambda(t))\nabla\eta(\textbf{x},\Lambda(t)),
\end{eqnarray}
under the prescription of a positive unit charge.
Noting the current density in the standard dynamics:
\begin{eqnarray}\label{stand-curren}
\textbf{j}(\textbf{x},t)&\equiv&
\textrm{Re}[\psi_0^{*}(\textbf{x},t)\frac{\hbar}{i m}\nabla
\psi_0(\textbf{x},t)]\nonumber\\
&=&\frac{\hbar}{m}\rho^2(\textbf{x},t)\nabla\eta(\textbf{x},t),
\end{eqnarray}
we find
\begin{eqnarray}\label{stand-ff-rel}
\textbf{j}_{FF}(\textbf{x},t)=
\alpha(t)\textbf{j}(\textbf{x}, \Lambda(t)).
\end{eqnarray}
Thus the standard current density becomes both squeezed and magnified by a time-scaling factor
$\alpha(t)$  in Eq. (\ref{nonunif-scale}) or Eq.(\ref{unif-scale}) as a result of the exact fast forwarding which enables acceleration of both amplitude and phase of the wave function throughout the time evolution.

Finally in this Section, we shall evaluate the expectation  of  energy of a particle in fast-forward dynamics and compare it with the corresponding expectation in standard dynamics.
We can formally rewrite the Schr\"{o}dinger equations in Eq.(\ref{MN-standeq}) and Eq.(\ref{ffeq}), respectively as
\begin{eqnarray}\label{tenergy1s}
i\hbar\frac{\partial\psi_{0}}{\partial t}=\hat{H}_0\psi_{0},
\end{eqnarray}
\begin{eqnarray}\label{tenergy1f}
i\hbar\frac{\partial\psi_{FF}}{\partial t}=\hat{H}_{FF}\psi_{FF},
\end{eqnarray}
where $\hat{H}_{0}$ and $\hat{H}_{FF}$ are taken as corresponding Hamiltonian operators.
We can write the expectation of energy in two cases as:
\begin{eqnarray}\label{tenergy2s}
\mathcal{E}_{0}(t)=\int\psi^{*}_{0}(\textbf{x},t)\hat{H}_{0}\psi_{0}(\textbf{x},t)d\textbf{x}
\end{eqnarray}
and
\begin{eqnarray}\label{tenergy2f}
\mathcal{E}_{FF}(t)=\int\psi^{*}_{FF}(\textbf{x},t)\hat{H}_{FF}\psi_{FF}(\textbf{x},t)d\textbf{x},
\end{eqnarray}
where the integration is over full space ($\textbf{x}\equiv (x,y,z)$).
Substituting Eqs.(\ref{tenergy1s}) and (\ref{tenergy1f}) into Eqs.(\ref{tenergy2s}) and (\ref{tenergy2f}) respectively, we obtain
\begin{eqnarray}\label{tenergy3s}
\mathcal{E}_{0}(t)=i\hbar\int\psi^{*}_{0}(\textbf{x},t)\frac{\partial\psi_{0}(\textbf{x},t)}{\partial t}d\textbf{x}
\end{eqnarray}
and
\begin{eqnarray}\label{tenergy3f}
\mathcal{E}_{FF}(t)&=&i\hbar\int\psi^{*}_{FF}(\textbf{x},t)\frac{\partial\psi_{FF}(\textbf{x},t)}{\partial t}d\textbf{x}\nonumber\\
&=&i\hbar\int\psi^{*}_{0}(\textbf{x},\Lambda)\frac{\partial\psi_{0}(\textbf{x},\Lambda)}{\partial\Lambda}\frac{\partial\Lambda}{\partial t}d\textbf{x}\nonumber\\
&=&i\hbar\alpha(t)\int\psi^{*}_{0}(\textbf{x},\Lambda)\frac{\partial\psi_{0}(\textbf{x},\Lambda)}{\partial\Lambda}d\textbf{x}.
\end{eqnarray}
$\alpha(t)$ comes from $\frac{\partial\Lambda}{\partial t}$ in Eq.(\ref{ff-t}).
Comparing Eq.(\ref{tenergy3f}) with Eq.(\ref{tenergy3s}), we have the relation between the expectations of energy between standard and fast-forward dynamics:
\begin{eqnarray}\label{tenergy4}
\mathcal{E}_{FF}(t)=\alpha(t)\mathcal{E}_{0}(\Lambda (t)),
\end{eqnarray}
which is similar to Eq.(\ref{stand-ff-rel}) and will play a vital role in the fast forward of quantum tunneling.

Now we shall apply the present scheme to several tunneling phenomena in quantum mechanics.
As for $\alpha(t)$, we shall choose a non-uniform factor in Eq.(\ref{nonunif-scale}) in the fast-forward time region, except when stated otherwise.

\renewcommand{\labelenumi}{\theenumi.}

\section{Fast forward of tunneling of wave packet dynamics}\label{FF-TUNNEL-WP}

Confining to the one-dimensional (1-d) motion, we now investigate the time evolution of a
localized wave packet when it runs through the delta-function
barrier. The initial wave packet centered at
$x=-x_{0}$ and having the width $\beta^{-1}$ and momentum $k$  is expressed as
\begin{eqnarray}\label{in-wp}
\psi^{(0)}(x,0)=\sqrt\beta e^{-\beta|x+x_{0}|}e^{ik(x+x_{0})}.
\end{eqnarray}
$\psi^{(0)}(x,0)$ satisfies the normalization condition
$\int_{-\infty}^\infty |\psi^{(0)}(x,0)|^2 dx=1$. Therefore, $<x>=-x_0$ and  $<p>=k$ at $t=0$.
While the wave function in Eq.(\ref{in-wp}) is non-differentiable at $x=-x_0$, it does not generate a discontinuity in
physical quantities like probability amplitude, current density, energy density, etc.

Time-dependent Schr\"odinger equation with delta
function barrier at $x=0$ is given by
\begin{eqnarray}\label{TDsch-barr}
[i\hbar\partial_{t}+(\hbar^{2}/2m)\partial^{2}_{x}]\psi_0(x,t)=V(x)\psi_0(x,t)
\end{eqnarray}
with $V(x)=V_{0}\delta(x)$.
In order to simplify the notation, we
shall use "natural unit" ($\hbar=m=1$) from now on.

The time evolution of $\psi_0$ for $t>0$  follows from
\begin{eqnarray}\label{method}
\psi_0(x,t)&=&\int_{-\infty}^\infty {\rm d}x' K_{0}(x,t|x',0)\psi^{(0)}(x',0) \nonumber\\
&-&V_{0}\int_{-\infty}^\infty\mathrm{d}x'
M(|x|+|x'|;-iV_{0};t)\psi_0(x',0).\nonumber\\
\end{eqnarray}
The first term on the r.h.s. of Eq.(\ref{method})
describes the time evolution of the free $(V_{0}=0)$ wave packet.
Here $K_{0}$  the free-particle propagator  given by
\begin{eqnarray}\label{A-propa}
K_{0}(x,t,|x',0)=\left(\frac{m}{2\pi i\hbar t}\right)^{1/2} \exp\left(i\frac{m(x-x')^{2}}{2\hbar t}\right).
\end{eqnarray}
$M(x;k;t)$ is  Moshinsky  function\cite{mos1,mos2}
defined in terms of the complementary error function
by
\begin{eqnarray}\label{A-Moshin}
M(x;k;t)=\frac{1}{2}e^{i(kx-k^{2}t/2)}\textrm{erfc}\left(\frac{x-kt}{\sqrt{2it}}\right),
\end{eqnarray}
which is interpreted as the wave function of a monochromatic particle that is confined to the left half- space $x \le 0$ at $t=0$.

The explicit solution $\psi_0(x,t)$ for $t>0$ was given by Elberfeld and Kleber\cite{ek} as
\begin{eqnarray}\label{wp-evol}
\psi_0(x,t)=\sqrt\beta[M(x+x_{0};k-i\beta;t)\nonumber\\+M(-x-x_{0};-k-i\beta;t)]\nonumber\\+V_{0}\sqrt\beta[S(x_{0},\lambda^{*};t)-S(x_{0},-\lambda;t)\nonumber\\+e^{-\lambda x_{0}}[S(0,-\lambda;t)+S(0,\lambda;t)]],
\end{eqnarray}
where $\lambda=\beta-ik$ and $S(\xi,\lambda;t)$ is defined by:
\begin{eqnarray}
S(\xi,\lambda;t)=[1/(V_{0}-\beta)][M(|x|+\xi;-iV_{0};t)\nonumber\\-M(|x|+\xi;-i\beta;t)].
\end{eqnarray}

The tunneling current evaluated just behind
the barrier at $x=0$ is:
\begin{eqnarray}\label{s-tun-curr}
j(+0,t)=\textrm{Im}[\psi_0^{*}(x,t)\partial_{x}\psi_0(x,t)]_{x=+0}  .
\end{eqnarray}
The current is  continuous, i.e,
$j(+0,t)=j(-0,t),$ owing to a nonabsorbing
potential barrier.
These are  results of the standard tunneling, i.e., tunneling dynamics on standard time scale.

Now we
analyze the fast forward of tunneling of the wave packet, and find the current density. By extracting the space-time dependent phase $\eta$ of the wave function
in Eq.(\ref{wp-evol}) in standard tunneling, one can obtain both vector and scalar potentials in Eqs.(\ref{aff-sol}) and (\ref{vff-sol}). Here $\eta$ is available only numerically because $\psi_0$ in Eq.(\ref{wp-evol}) is  a linear combination of special functions.  Under these driving potentials, one can generate
the fast-forward state of tunneling of the wave packet through the
barrier:
\begin{eqnarray}\label{psiFF}
\psi_{FF}(x,t)&\equiv &\psi_{0}(x,\Lambda(t))\nonumber\\
&=&\sqrt\beta[M(x+x_{0};k-i\beta;\Lambda(t))\nonumber\\
&+&M(-x-x_{0};-k-i\beta;\Lambda(t))]\nonumber\\
&+&V_{0}\sqrt\beta[S(x_{0},\lambda^{*};\Lambda(t))-S(x_{0},-\lambda;\Lambda(t))\nonumber\\
&+&e^{-\lambda x_{0}}[S(0,-\lambda;\Lambda(t))+S(0,\lambda;\Lambda(t))]],\nonumber\\
\end{eqnarray}
which accelerates both amplitude and phase of Eq.(\ref{wp-evol}) exactly.
It should be noted that, without having recourse to $\eta$, $\psi_{FF}$ in Eq.(\ref{psiFF})
is obtained from $\psi_0$ by the definition itself in Eq.(\ref{ffwf}).
From Eq.(\ref{stand-ff-rel}), the current density evaluated at $x=+0$
for the fast-forward 1-d tunneling phenomenon is:
\begin{eqnarray}\label{ff-tun-curr}
j_{FF}(+0,t)=\alpha(t) j(+0,\Lambda(t)).
\end{eqnarray}

In our numerical analysis
in Sections \ref{FF-TUNNEL-WP} and \ref{Moshinsky}, we shall use typical space and time units like
$L=10^{-2} \times$ {\it the linear dimension of a device} and $\tau=10^{-2} \times$
{\it the phase coherent time}, besides the natural unit ($\hbar=m=1$).  Then, any length, wave number and time are scaled by $L$, $L^{-1}$ and $\tau$, respectively. In this scaling, we choose $x_0=2, k=2$ and $\beta=1$.
We shall show the standard dynamics up to the standard final time $T=2.5$ and its fast-forward version up to
the shortened final time $T_{FF}\equiv\frac{T}{\bar{\alpha}}=0.5$ with  the
mean acceleration factor $\bar{\alpha}=5$.
\begin{figure}[htb]
\centering
\includegraphics[width=0.8\linewidth]{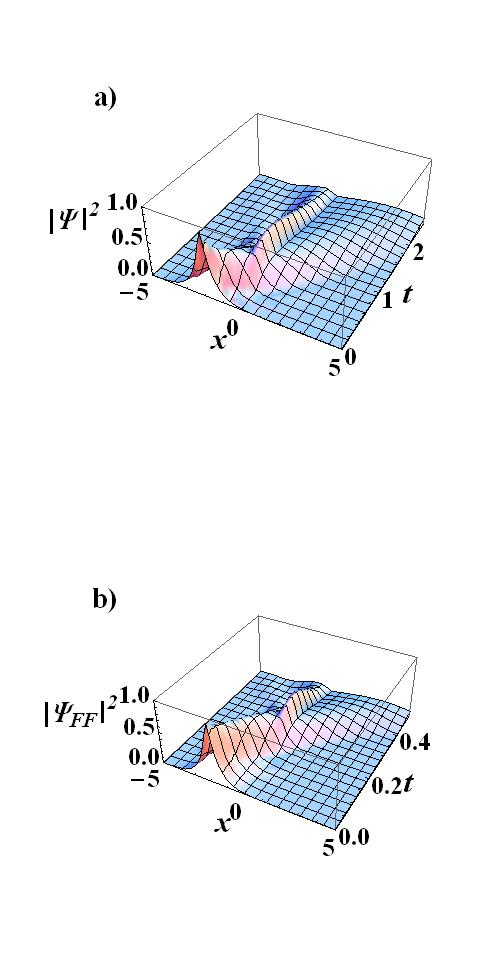}
\caption{(Color online)  Three-dimensional (3d) plot of the
density distribution
for the tunneling wave function
starting from the exponential wave packet in Eq.(\ref{in-wp}). In the scaling described below Eq.(\ref{ff-tun-curr}), $x_0=2, k=2, \beta=1, T=2.5, T_{FF}=0.5$
and $\bar{\alpha}=5$.  The barrier height here is $V_0=1$: (a) standard dynamics given
by $|\psi_0 (x,t)|^{2}$ for $0 \le t \le T$;
(b) fast-forward dynamics given by $|\psi(x,\Lambda(t))|^{2}$ for $0 \le t \le T_{FF}$.
In Figs. \ref{wp-tunnel-current}-\ref{asymp-curr}, the same scaling as in this figure is used. }
\label{bifurc-wf}
\end{figure}

In Fig. \ref{bifurc-wf}, we see the exponential wave function partly go through the barrier and is partly reflected back after its collision with the barrier. The dynamics up to $T$ on the standard time scale is reproduced in the fast-forward dynamics up to
$T_{FF}$: The phenomena in the latter is just the squeezing (along time axis) of the one in the former. The minor discrepancy in the similarity of figures between upper and lower panels in time axis is due to the non-uniform time-scaling factor in Eq.(\ref{nonunif-scale}), while the wave function amplitude w.r.t. to space axis is exactly reproduced at each time of the fast forward dynamics.

\begin{figure}[htb]
\centering
\includegraphics[width=0.9\linewidth]{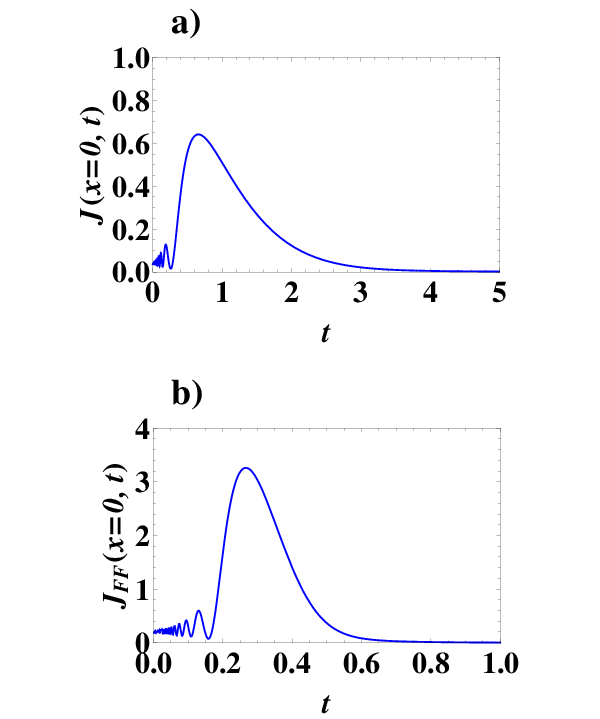}
\caption{(Color online) Temporal behavior of tunneling current density
at the position just behind the barrier ($x=+0$) for the tunneling wave function
starting from the exponential wave packet in Eq.(\ref{in-wp}). $T=5, T_{FF}=1$ and $\bar{\alpha}=5$. The barrier height is $V_0=1$: a) $j(t)$; b) $j_{FF}(t)$. Note: scales of the horizontal and vertical axes differ between upper and lower panels.}
\label{wp-tunnel-current}
\end{figure}

Figure \ref{wp-tunnel-current} shows the tunneling current at the position just behind the barrier ($x=+0$) as a function of time $t$. Here we choose $T=5, T_{FF}=1$ and $\bar{\alpha}=5$. We find the the temporal behavior of the current $j$ in the standard tunneling is both squeezed and amplified in that of the current $j_{FF}$ in the fast-forward tunneling.

\begin{figure}[htb]
\centering
\includegraphics[width=0.8\linewidth]{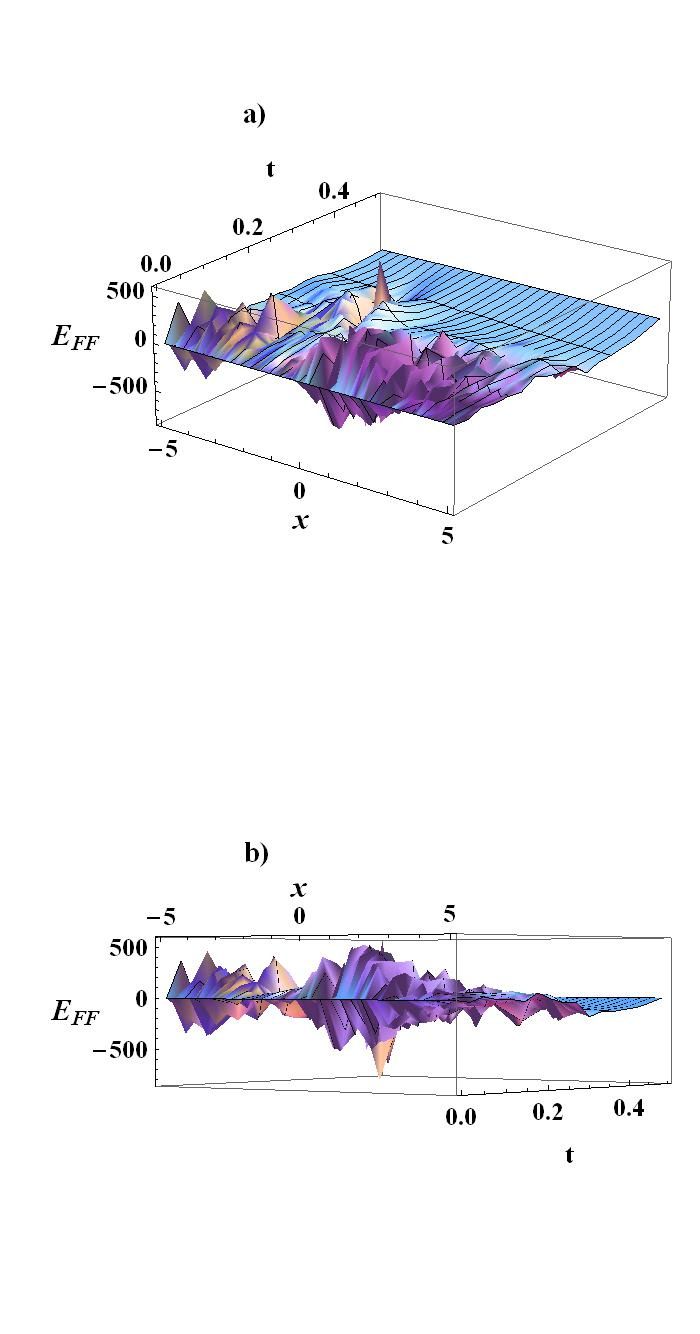}
\caption{(Color online)  3-d plot of Electric field $E_{FF}$
as a function of $x$ and $t$: a) top view; b) side view. }
\label{tunAFVF}
\end{figure}

Let us define the tunneling rate (: tunneling power) as
\begin{eqnarray}\label{tunnel-rate}
\Gamma=\frac{\int_{0}^{T} j(x=+\epsilon,t)dt}{T},
\end{eqnarray}
where $T$ here means the final time that the tunneling phenomenon is almost completed.
The corresponding rate for the fast-forwarded case is:
\begin{eqnarray}\label{ff-tunnel-rate}
\Gamma_{FF}=\frac{\int_{0}^{T_{FF}} j_{FF}(x=+\epsilon,t)dt}{T_{FF}}.
\end{eqnarray}
In Fig.\ref{wp-tunnel-current} we see $T=5$ and $T_{FF}=1$ in the standard and fast-forward tunnelings, respectively.
Noting Eq.(\ref{ff-tun-curr}), we find the numerator of r.h.s of Eq.(\ref{ff-tunnel-rate}) is equal to that of Eq.(\ref{tunnel-rate}). Then one can conclude:
\begin{eqnarray}\label{ratio}
\Gamma_{FF}=\frac{T}{T_{FF}}\Gamma=\bar{\alpha}\Gamma.
\end{eqnarray}
Thus the fast forwarding lets the standard tunneling rate enhanced by a factor of the mean magnification time scale.  This is a great advantage of the fast forward of quantum tunneling.

Figure \ref{tunAFVF}  shows a 1-d version of the driving electric field $\textbf{E}_{FF}$ 
in Eq.(\ref{ele-fld}) to generate the exact fast forward of the tunneling dynamics, which is given by
\begin{eqnarray}\label{ele-fld-1d}
E_{FF}&=&\hbar \dot{\alpha}\partial_x \eta+\hbar \frac{\alpha^2-1}{\alpha}\partial_t\partial_x \eta \nonumber\\
&+&\frac{\hbar^{2}}{m}(\alpha^2-1)\partial_x \eta \cdot \partial_x ^2 \eta.
\end{eqnarray}
The bottom panel of Figure \ref{tunAFVF}(c) shows a 3-d plot of $E_{FF}$ on $x-t$ plane, while the top and middle panels are its cross sections in $t$ and $x$ directions, respectively.
In SI unit for electric field, our dimensionless $E_{FF}$ corresponds to
$E^{FF}_{SI}=\frac{m_e c \omega}{e}\times E_{FF}\sim\frac{10^6}{\lambda} E_{FF}$ where
$m_e, e, c, \omega$ and $\lambda$ are electron mass, electron charge, velocity of light, frequency of laser light and its wave length, respectively. Typical value $E_{FF}=100$ in ordinates in Fig. \ref{tunAFVF}
in case of IR lasers of wave length $\sim$ 1$\mu$m means $E^{FF}_{SI}=10^{14}$. The
driving electric field shown in Fig. \ref{tunAFVF}  can be implemented
using for instance
a rapidly moving laser beam to create a possibly dynamic
time-averaged optical dipole potential\cite{Hend}.

Before closing this Section, we should note: the standard tunneling here is autonomous (i.e., $V_0= const.$) and thereby the total energy of the electron $\mathcal{E}_0=const.$, while the corresponding fast-forward tunneling is non-autonomous with
the time-dependent total energy $\mathcal{E}_{FF}(t)$.  $\mathcal{E}_{FF}(t)$ satisfies $\mathcal{E}_{FF}(t)=\alpha(t)\mathcal{E}_{0}$ as a special case of
Eq.(\ref{tenergy4}).
In the tunneling phenomenon of standard dynamics,
we always see the inequality $\mathcal{E}_0 < V_0$(barrier height).
Then we can define $\alpha_{max} \equiv \frac{V_0}{\mathcal{E}_0} (>1)$
and choose the time scaling factor $\alpha (t)$ as
\begin{eqnarray}\label{amax}
1\leq \alpha (t) < \alpha_{max},
\end{eqnarray}
which guarantees the inequality $\mathcal{E}_{FF}(t)=\alpha(t)\mathcal{E}_0 < V_0$.
In conclusion, so long as Eq.(\ref{amax}) is satisfied,
the fast-forwarded dynamics here is also the tunneling phenomenon keeping the particle's energy $\mathcal{E}_{FF}$ below the barrier height $V_0$ throughout the time evolution and the time scaling works more effectively for the particle with lower incident energy.
The same assertion as above  will hold in the following Sections.

 \section{Fast forward tunneling dynamics from Moshinsky shutter}\label{Moshinsky}
Now let's  investigate the dynamics of a
monochromatic beam of noninteracting particles of mass $m$ and
energy $\hbar^{2}k^{2}/2m$ moving parallel to the $x$ axis from the
left to the right. Until $t<0$,  the beam is assumed being stopped by the shutter at $x=0$
perpendicular to the beam. If at $t=0$ the shutter is opened,
the transient particle current is observed  at a distance $x$
from the shutter. This problem was first solved by Moshinsky\cite{mos1}, and then received a renewed attention  by Elberfeld and Kleber\cite{ek}, who introduced a delta-function barrier with a finite height at $x=0$ and considered the tunneling through it.

The shutter acts as a perfect
absorber. Then, the wave function that represents a
particle of the beam is initially given by
\begin{eqnarray}\label{in-step}
\psi^{(0)}(x,t=0)=\Theta(-x)e^{ikx}
\end{eqnarray}
with
the step function $\Theta(x)=0$ and $1$ for $x<0$ and  for $x>0$, respectively.
The time-dependent Schr\"odinger equation with a delta
function barrier is the same as in Eq.(\ref{TDsch-barr}).
By applying the same method as in Eq.(\ref{method}), the solution satisfying the initial condition in Eq.(\ref{in-step}) was obtained \cite{ek} as:
\begin{eqnarray}\label{st-tun-shutt}
\psi_0(x,t)=M(x;k;t)+[V_{0}/(V_{0}-ik)]\nonumber\\\times [M(|x|; -iV_{0};t)-M(|x|;k;t)].
\end{eqnarray}
The tunneling current just behind the barrier is evaluated using Eq.(\ref{s-tun-curr}).

\textbf{\textbf{A}. Fast forward of Moshinsky shutter in the presence of
delta-function barrier}

\begin{figure}[htb]
\centering
\includegraphics[width=0.8\linewidth]{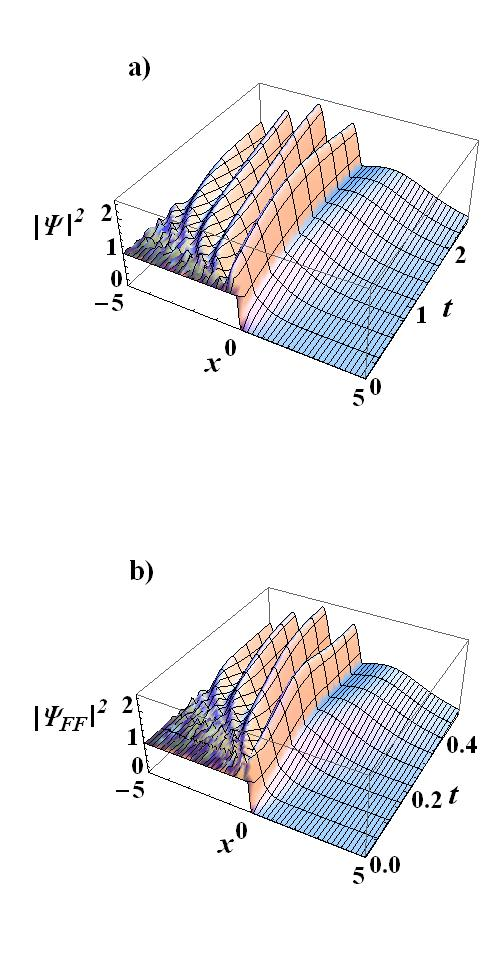}
\caption{(Color online) 3-d plot of the density distribution
 (vertical direction)
as a function of $x$ and $t$. The density distribution
illustrates the tunneling dynamics of the semi-infinite
wave train ($k=2$) penetrating through the
delta-barrier with the height $V_0=1$ located at $x=+0$: (a) standard tunneling dynamics until the final time $T=2.5$;
(b) fast-forward tunneling dynamics until $T_{FF}=0.5$ with the average time scaling factor
$\bar{\alpha}=5$.}
\label{3d-mosh-tun}
\end{figure}

\begin{figure}
\centering
\includegraphics[width=0.9\linewidth]{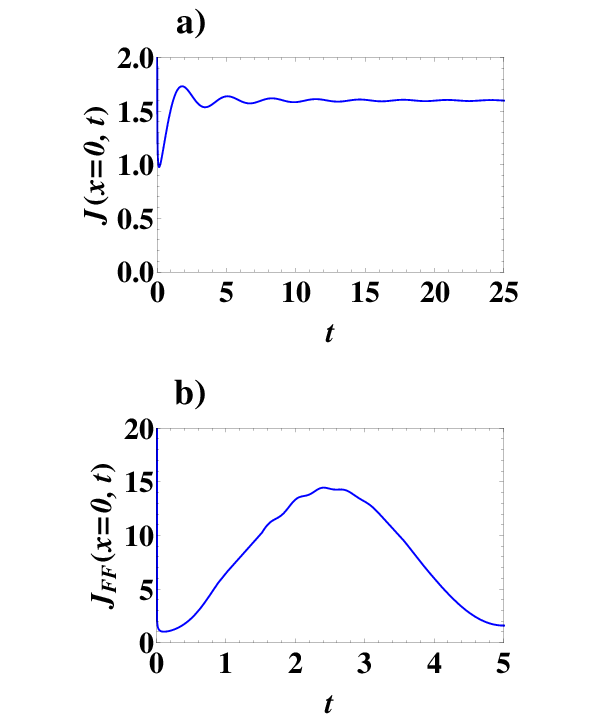}
\caption{(Color online)  Tunneling current densities  for an initial
semi-infinite wave train in the case of  $k=2, V_0=1$ and
$\bar{\alpha}=5$: a) $j(+0,t)$ with $T=25$;  b) $j_{FF}(+0,t)$ with
$T_{FF}=\frac{T}{\alpha}=5$.  Note: scales of the horizontal and vertical axes
differ between upper and lower panels.} \label{mosh-tun-curr}
\end{figure}

\begin{figure}
\centering
\includegraphics[width=0.8\linewidth]{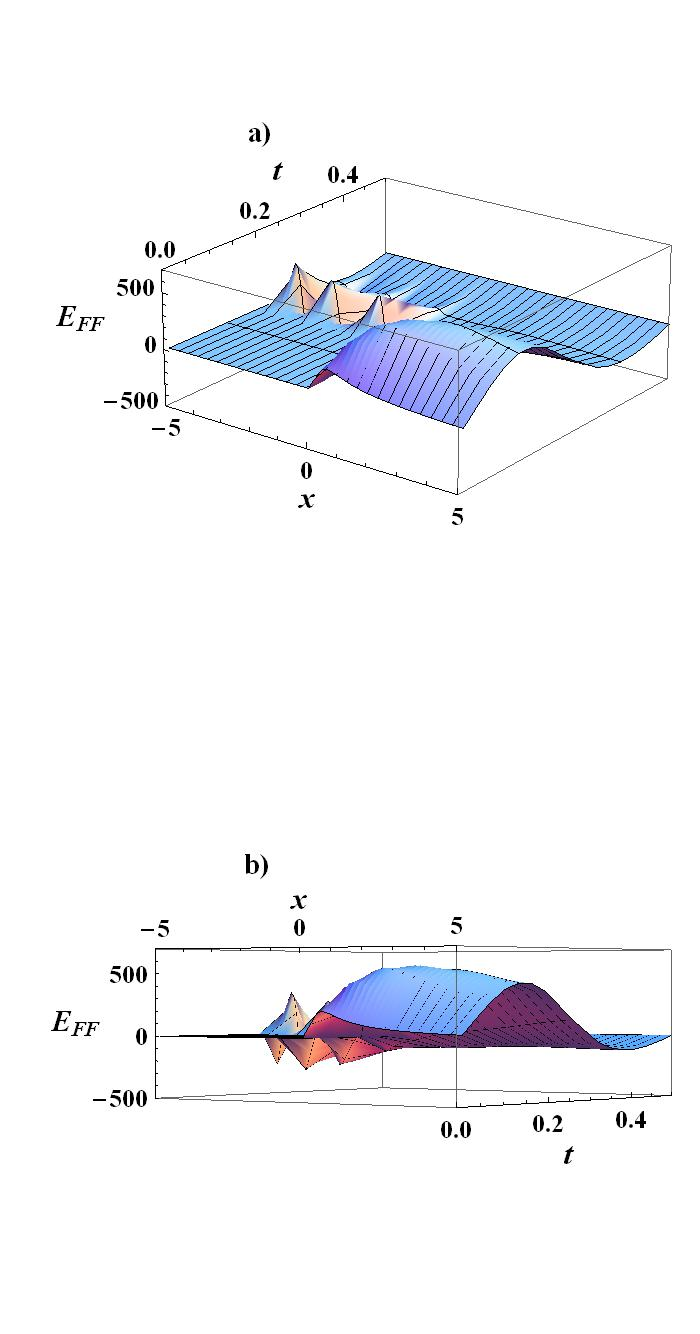}
\caption{(Color online)  3-d plot of electric field $E_{FF}$ as a function of $x$ and $t$:  a) top view; b) side view. }
\label{mosh-vff-aff}
\end{figure}

We now analyze the fast forwarded tunneling for
Moshinsky shutter. One can evaluate the phase $\eta$ of the wave function solution in Eq.(\ref{st-tun-shutt}), which is used to determine both driving vector and scalar potentials in Eqs.(\ref{aff-sol}) and (\ref{vff-sol}). By applying these driving potentials, we obtain the exact
fast forwarded state, which is given by replacing $t$ by $\Lambda(t)$ in Eq.(\ref{ff-t}) as:
\begin{eqnarray}\label{ff-mosh-wave}
\psi_{FF}(x,t)&\equiv& \psi_{0}(x,\Lambda(t))\nonumber\\
&=&M(x;k;\Lambda(t))+[V_{0}/(V_{0}-ik)]\nonumber\\
&\times& [M(|x|; -iV_{0};\Lambda(t))-M(|x|;k;\Lambda(t))]. \nonumber\\
\end{eqnarray}
Concerning the relation between $\psi_{FF}$ and $\eta$, one should recall the notion
just before and after Eq.(\ref{psiFF}).
The fast-forwarded tunneling current  is given by Eq.(\ref{ff-tun-curr}).

Using the same units as described below Eq.(\ref{ff-tun-curr}), we choose $k=2, V_0=1, \bar{\alpha}=5$, $T=2.5$ and $T_{FF}=0.5$ and show in Fig.\ref{3d-mosh-tun} the
density profiles of wave functions in both cases of  the  standard and fast-forward (or accelerated) dynamics.
We find that the time evolution of $|\psi_{0}(x,t)|^{2}$ is squeezed in $|\psi_{FF}(x,t)|^{2}$ by the factor
$\frac{1}{\bar{\alpha}}$. Interference between the incoming and reflected
waves leads to an oscillatory profile for $x<0$. For $x>0$ the outgoing wave shows a
rather smoothly-varying density profile.

In the case of $V_0=1, \bar{\alpha}=5$, we show in Fig.\ref{mosh-tun-curr} the tunneling current just behind the barrier in wider time range with use of $T=25$ and $T_{FF}=5$. The temporal behavior of the current $j(+0,t)$
in the standard dynamics, which shows a  very slow decrease w.r.t. $t$,  is squeezed and enhanced in that of $j_{FF}(+0,t)$ in the fast-forward dynamics.
As noted in the previous Section, Eq.(\ref{ff-tun-curr}) leads to the equality:
\begin{eqnarray}\label{integrate}
\int_{0}^{T_{FF}} j_{FF}(x=+0,t)dt=\int_{0}^{T} j(x=+0,t)dt.
\end{eqnarray}
In case of Fig.\ref{mosh-tun-curr}, a rough estimate for the r.h.s of Eq.(\ref{integrate}) is $25\times 1.6$, which agrees with  $5 \times\bar {J}_{FF}$(: average fast-forward current)  evaluated for l.h.s. Therefore $\bar {J}_{FF}
\sim 8$, as seen in Fig.\ref{mosh-tun-curr}. In general, the tunneling current in the standard dynamics greatly decreases  when the barrier height $V_0$ becomes much larger than unity. But, a suitable  fast-forward mechanism recovers the current for the case of $V_0=1$, which will be described by asymptotic argument in the next Subsection.

The fast forward state in Eq.(\ref{ff-mosh-wave}) can be generated as a solution of the time-dependent Schr\"{o}dinger equation in Eq.(\ref{ffeq}) for the charged particle in the presence of vector $A_{FF}(x,t)$ and scalar
$V_{FF}(x,t)$ potentials. Figure \ref{mosh-vff-aff} shows the corresponding electric field $E_{FF}$ evaluated by Eq.(\ref{ele-fld-1d})  as a function of $x$ and $t$.

\textbf{\textbf{B}. Asymptotic approach in case of  a very high barrier}

In the limit of a very high barrier, the tunneling current becomes negligibly small. But by a suitable choice of the time scaling $\alpha(t)$, one can recover the standard magnitude of the tunneling current, which we shall show below.

We first rewrite the wave function solution on the r.h.s. of the barrier (i.e. for $x>0$) in Eq.(\ref{st-tun-shutt}) as
\begin{eqnarray}\label{st-tun-shutt2}
\psi_0(x,t)=-\frac{ik}{V_{0}-ik}M_1+\frac{V_{0}}{V_{0}-ik}M_2.
\end{eqnarray}
Here
\begin{eqnarray}\label{st-tun-reor}
M_1\equiv M(x;k;t),\qquad M_2\equiv M(x;-iV_0;t),
\end{eqnarray}
where  $M(x;k;t)$ is the Moshinsky function defined in Eq.(\ref{A-Moshin}). With use of a new variable
\begin{eqnarray}\label{new-var-z}
z=\frac{x+iV_0t}{\sqrt{2t}}e^{-i\frac{\pi}{4}},
\end{eqnarray}
we see $(-iV_0)x-\frac{(-iV_0)^2t}{2}=-iz^2+\frac{x^2}{2t}$
and
\begin{eqnarray}\label{M2asym}
M_2=\frac{1}{2}e^{z^2+\frac{x^2}{2t}i}{\rm erfc}(z).
\end{eqnarray}

We shall now concentrate on the asymptotic region given by
$V_0 \gg 1$ with $k=O(1), t=O(1)$ and $x \ll 1$, which leads to $|z| \gg 1$.
Then we find
${\rm erfc}(z) \sim \frac{e^{-z^2}}{\sqrt{\pi}z}$ and
\begin{eqnarray}\label{M2asym2}
M_2\sim \frac{1}{2\sqrt{\pi}z}e^{\frac{x^2}{2t}i}.
\end{eqnarray}
Noting $z|_{x=0}=V_0\sqrt{\frac{t}{2}}e^{i\frac{\pi}{4}}$ and $\frac{\partial z}{\partial x}|_{x=0}=\frac{1}{\sqrt{2t}}e^{-i\frac{\pi}{4}}$,
we have $M_2|_{x=0}= \frac{1}{V_0\sqrt{2\pi t}}e^{-i\frac{\pi}{4}}$
and $\frac{\partial M_2}{\partial x}|_{x=0}=\frac{1}{V_0^2 \sqrt{2\pi t^3}}e^{i\frac{\pi}{4}}$, which results in
\begin{eqnarray}\label{M2prod}
M_2^*\frac{\partial M_2}{\partial x}|_{x=0}=\frac{i}{2\pi t^2V_0^3}(1+O(\frac{1}{V_0^2})).
\end{eqnarray}
In a similar way, we find:
\begin{eqnarray}\label{M12prod}
M_1^*\frac{\partial M_2}{\partial x}|_{x=0}&=&M^*(0;k;t)\frac{1}{\sqrt{2\pi t^3}V_0^2}e^{i\frac{\pi}{4}}(1+O(\frac{1}{V_0^2})),\nonumber\\
M_2^*\frac{\partial M_1}{\partial x}|_{x=0}&=&\frac{1}{\sqrt{2\pi t}V_0}e^{i\frac{\pi}{4}}\frac{\partial M(x;k;t)}{\partial x}|_{x=0}(1+O(\frac{1}{V_0^2})).\nonumber\\
\end{eqnarray}

\begin{figure}
\centering
\includegraphics[width=0.8\linewidth]{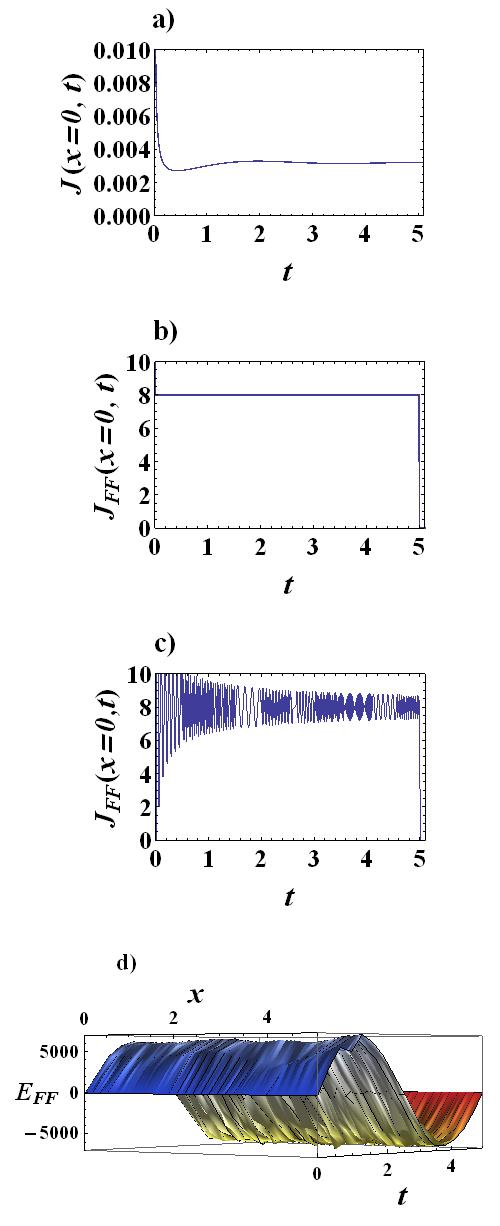}
\caption{(Color online) Standard and fast-forwarded tunneling current of a
semi-infinite wave train in case of
$V_{0}=50$ and  $k=2$.  In the fast-forwarded case, the uniform time scaling with $\bar{\alpha}=V_{0}^2=2500$ is employed during the period between $0$ and $T_{FF}$ with
$T_{FF}=\frac{T}{\bar{\alpha}}=\frac{12500}{2500}=5$. Time range depicted is $0\le t \le 5$. (a) Standard current available from Eqs. (\ref{s-tun-curr}) and (\ref{st-tun-shutt});  (b)  Fast-forwarded exact current in
Eqs. (\ref{ff-tun-curr}) and (\ref{ff-mosh-wave}); (c) Fast-forwarded asymptotic current in Eq.(\ref{asym-uff-cur}). The rapid oscillation in the fast-forward time region comes from the 2nd term in the last expression of Eq.(\ref{asym-uff-cur}) which includes an un-cancelled factor $e^{-ik^2\Lambda (t) /2}$ of Moshinsky function in Eq.(\ref{A-Moshin}); d) 3-d plot of the driving electric field to realize the fast-forwarded exact current in case of b).}
\label{asymp-curr}
\end{figure}

Using the decomposition in Eq.(\ref{st-tun-shutt2}), the standard current just behind the barrier can
be expressed as
\begin{eqnarray}\label{ext-curr}
j(x=+0,t)&=&\textrm{Im}\left[\psi^{\ast}(x=+0,t)\frac{\partial\psi}{\partial x}\big|_{x=+0}\right]
\nonumber\\
&=&\frac{k^{2}}{V^{2}_{0}+k^{2}}\textrm{Im}(M_{1}^{\ast}\partial_{x}M_{1})|_{x=0}\nonumber\\
&+&\frac{kV_{0}}{V^{2}_{0}+k^{2}}\textrm{Re}(M_{1}^{\ast}\partial_{x}M_{2})|_{x=0}\nonumber\\
&+&\frac{kV_{0}}{V^{2}_{0}+k^{2}}\textrm{Re}(M_{2}^{\ast}\partial_{x}M_{1})|_{x=0}\nonumber\\
&+&\frac{V_0^{2}}{V^{2}_{0}+k^{2}}\textrm{Im}(M_{2}^{\ast}\partial_{x}M_{2})|_{x=0}.\nonumber\\
\end{eqnarray}
Noting the asymptotics in Eqs.(\ref{M2prod}) and (\ref{M12prod}),
one sees that the first and third terms give dominant contributions of $O(\frac{1}{V_0^2})$ and other terms give minor contributions of $O(\frac{1}{V_0^3})$. Then
$j(x=+0,t)$ becomes asymptotically:
\begin{eqnarray}\label{curr-bigV0}
j(x=+0,t)
&=&\frac{k^2}{V_0^{2}}\textrm{Im}(M^{\ast}(0;k;t)\partial_{x}M(x;k;t)|_{x=0}\nonumber\\
&+&\frac{k}{V_0^{2}}\textrm{Re}(\frac{1}{\sqrt{2\pi t}}e^{i\frac{\pi}{4}}\partial_{x}M(x;k;t)|_{x=0}) \nonumber\\
&+&O(\frac{1}{V_0^3}).
\end{eqnarray}
The tunneling current in the standard dynamics has proved to be of $O(\frac{1}{V_0^2})$,
which is very small.
However, the idea of fast forward can recover the current in the case of $V_0=O(1)$.
In fact, by applying the driving vector and scalar  potentials in Eqs.(\ref{aff-sol}) and (\ref{vff-sol}), we can realize the exact fast-forward state in the time domain $0 < t <T_{FF}$ with $T_{FF}$ in
Eq.(\ref{T-Tff}), and its corresponding fast-forward current is given by Eq.(\ref{ff-tun-curr}).
Therefore, if we shall use a large enough magnification time-scaling factor $\alpha (t)$ with its mean value $\bar{\alpha}=O(V_0^2)$,
the current in the case of the small barrier ($V_0=O(1)$) will be recovered.

To make the quantitative argument, let's employ the uniform scaling factor in Eq.(\ref{unif-scale}).
Then the fast-forward current is given by
\begin{eqnarray}\label{asym-uff-cur}
j_{FF}(x&=&+0,t)=\bar{\alpha}j(x=+0,\Lambda (t))\nonumber\\
&=&\frac{k^2\bar{\alpha}}{V_0^{2}}\textrm{Im}(M^{\ast}(0;k;\Lambda (t))\partial_{x}M(x;k;\Lambda (t))|_{x=0}\nonumber\\
&+&\frac{k\bar{\alpha}}{V_0^{2}}\textrm{Re}(\frac{1}{\sqrt{2\pi \Lambda (t)}}e^{i\frac{\pi}{4}}\partial_{x}M(x;k;\Lambda (t))|_{x=0}). \nonumber\\
\end{eqnarray}
Noting $\Lambda (t)=O(1)$ in the fast-forward time domain, Eq.(\ref{asym-uff-cur}) shows that if we shall choose
\begin{eqnarray}\label{recov}
\bar{\alpha}=V_0^2,
\end{eqnarray}
the tunneling current in the high barrier case ($V_0\gg 1$) will recover the value in the low barrier case  ($V_0=O(1)$).

Figure \ref{asymp-curr} shows: The fast forwarding with use of the driving electric field makes a  negligible tunneling current for the case of a very high barrier with $V_0 \gg 1$  increased to the value for the case of a standard barrier with $V_0 =O(1)$.  In fact, in case of $V_0=50$, $j=O(10^{-3})$ in Fig.\ref{asymp-curr}(a), but $j_{FF}=O(10)$ in Figs. \ref{asymp-curr}(b), (c) in the fast-forward time region $0\le t \le T_{FF}(=5)$.

\section{Fast forward of macroscopic tunneling}\label{macro}
\begin{figure}
\centering
\includegraphics[width=0.6\linewidth]{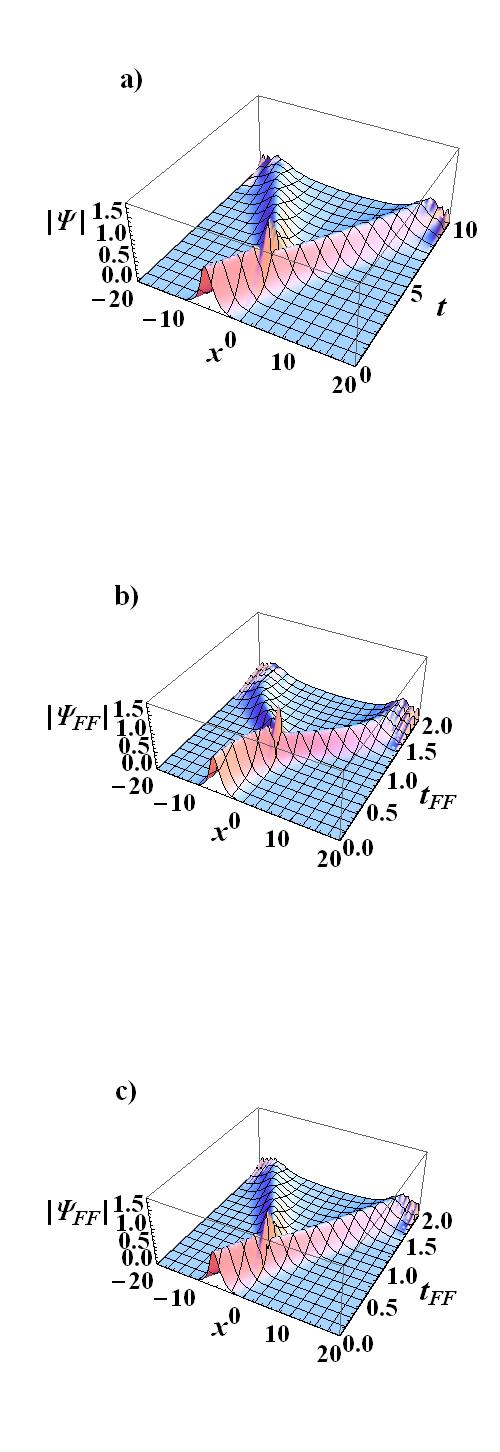}
\caption{(Color online) 3-d plot of
$|\psi(x,t)|$ (vertical direction)
as a function of $x$ and $t$:  (a) Standard tunneling dynamics of the soliton where $\psi_0(x,t)$ satisfies
Eq.(\ref{sol-standeq}) with $V_{0}=30, v=2.25, x_{0}=6$;
(b)  Fast-forward  tunneling dynamics of the soliton where $\psi_{FF}(x,t)$ satisfies
Eq.(\ref{ffsol-split}) under the non-uniform time-scaling factor $\alpha(t)$
with its mean $\bar{\alpha}=5$
in Eq.(\ref{nonunif-scale}).
$V_{0}, v, x_{0}$ are the same as in Fig.\ref{split-barrless}(a); (c) The same as in
Fig.\ref{split-barrless}(b) except for the uniform time scaling $\alpha(t)=\bar{\alpha}=5$.
In Figs. \ref{BEC-curr} and \ref{sol-drive}, the same space and time units as in this figure are used.}
\label{split-barrless}
\end{figure}

The theory of fast-forward can also be applied to the macroscopic
quantum mechanics. We shall consider the fast-forwarded tunneling of a solitonic wave packet in
1-d Bose-Einstein condensates (BEC) governed by nonlinear Schr\"odinger
equation in Eq.(\ref{MN-standeq}) with the  barrier at origin,
$V(x)=V_0\delta(x)$. The standard dynamics for $\psi_{0}$ is described by:
\begin{eqnarray}\label{sol-standeq}
i\hbar\partial_t \psi_{0}=-\frac{\hbar^2}{2m}\partial_x^2\psi_{0}+V_0\delta(x)\psi_{0}
-c_0|\psi_{0}|^{2}\psi_{0}.
\end{eqnarray}
On the other hand, the governing equation  for the fast-forward function $\psi_{FF}$ is given by (see the notice just below Eq.(\ref{ffeq})):
\begin{eqnarray}\label{ffsol-split}
\imath\hbar\partial_t{\psi_{FF}}&=&
\left(\frac{1}{2m}(\frac{\hbar}{i}\partial_x - A_{FF})^2+
V_{FF}+V_0\delta(x)\right)\psi_{FF}\nonumber\\
&-& c_0|\psi_{FF}|^2\psi_{FF}.
\end{eqnarray}

\begin{figure}
\centering
\includegraphics[width=0.9\linewidth]{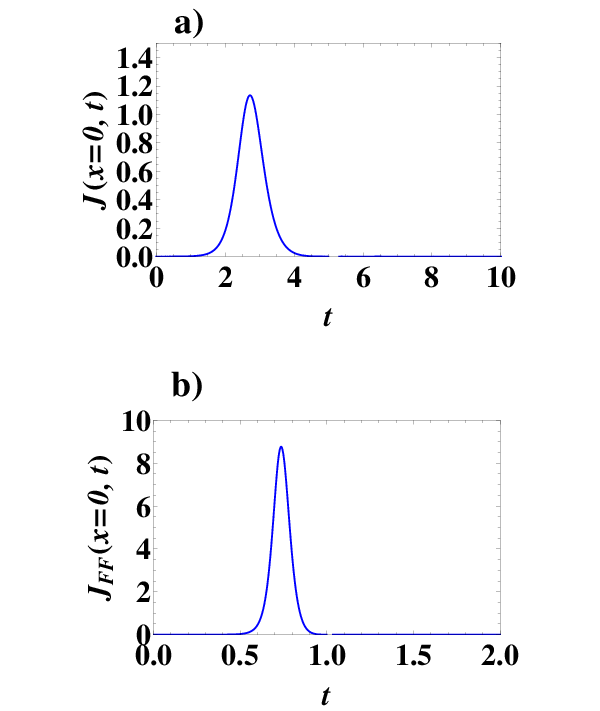}
\caption{(Color online) Tunneling current densities at $x=+0$:
(a) Standard tunneling current;
(b) Fast-forward  tunneling current. Note: scales of the horizontal and vertical axes differ between upper and lower panels.}
\label{BEC-curr}
\end{figure}

Below, besides the natural unit ($\hbar=m=1$) we shall employ the same units as used in the previous Sections on microscopic quantum dynamics.  Namely, space and time are scaled by $L=10^{-2} \times$ {\it the system size} and $\tau=10^{-2}\times$ {\it the dissipation time}, respectively, and we put the nonlinearity constant
$c_0$ (scaled by $L\tau^{-1}$)$=1$. Then Eqs. (\ref{sol-standeq}) and (\ref{ffsol-split}) become dimensionless, which we shall analyze.
If there is no barrier, the solution of Eqs.(\ref{sol-standeq}) is a travelling Zakharov-Shabat's soliton \cite{zakh}
given by
\begin{eqnarray}
\psi^{(0)}(x,t)=A {\rm sech}[(A(x-vt)]e^{i\phi_0+ivx + i(A^2-v^2)t/2}
\end{eqnarray}
with $A$ and $v$ for the amplitude and propagation velocity, respectively.
\begin{figure}[htb]
\centering
\includegraphics[width=0.9\linewidth]{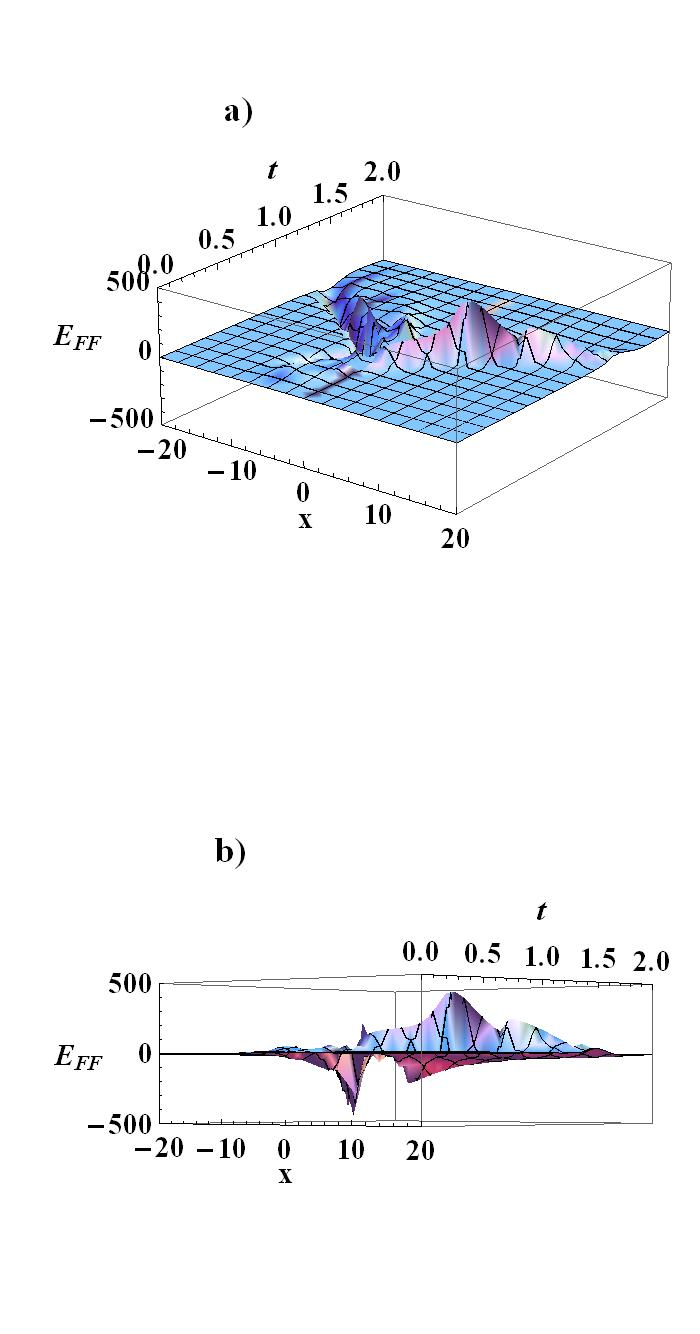}
\caption{ (Color online)  3-d
plot of the electric field $E_{FF}$ as a function of $x$ and $t$:
a) top view; b) side view.} \label{sol-drive}
\end{figure}

In the presence of the barrier, we numerically solve Eq.(\ref{sol-standeq}) with use of an initial profile:
\begin{eqnarray}
\psi^{(0)}(x,t)= {\rm sech}(x+x_0)e^{ivx},
\end{eqnarray}
which stands for the soliton with $A=1$ and initial position
$x=-x_0 (x_0 \gg 1)$ for center of mass.

In Fig.\ref{split-barrless}(a), we show the amplitude $|\psi(x,t)|$ of the soliton  as a function of $x$ and $t$ in standard time.
The soliton located at $x=-x_0$ moves to the right and after the time
$t=\frac{x_0}{v}$ that it
collides with the barrier at $x=0$. Then it splits into two
parts: reflected and transmitted ones which are moving to left
and right, respectively. The result accords with the one by Holms et al.\cite{holm}.  Increase of the barrier height $(V_0)$ diminishes the transmitted part, namely decreases the tunneling rate.

According to the idea of the fast forward, the same wave function patterns as seen in the time domain $0<t<T$ can be realized
in the shortened time domain $0<t<T_{FF}$ (see Eq.(\ref{T-Tff})) with use of  $A_{FF}$ and $V_{FF}$ in Eq.(\ref{aff-sol}) and Eq.(\ref{vff-sol}), respectively.

Using the non-uniform time-scaling factor $\alpha (t)$ with its mean $\bar{\alpha}=5$ in Eq.(\ref{nonunif-scale}), we have solved Eq.(\ref{ffsol-split}). In Fig.\ref{split-barrless}(b), the amplitude $|\psi_{FF}(x,t)|$ of the soliton is shown. At the shortened time
$t \sim \frac{x_0}{\bar{\alpha}v}$ the soliton collides with the delta-barrier at $x=0$ and the splitting process is also shortened.
In Fig.\ref{split-barrless}(c), we show $|\psi_{FF}(x,t)|$ by solving Eq.(\ref{ffsol-split}) with use of the uniform scaling factor $\bar{\alpha}=5$ in Eq.(\ref{unif-scale}). In this case we can see
the exact time-squeezed version of soliton dynamics in Fig.\ref{split-barrless}(a).
The soliton reaches the barrier at $t = \frac{x_0}{\bar{\alpha}v}$ and transmitted and reflected patterns are shortened by the constant time scaling $\bar{\alpha}$.

Now we shall compute the tunneling current $j(+0,t)$ and $j_{FF}(+0,t)$ at $x=+0$ in Eqs.(\ref{s-tun-curr})  and (\ref{ff-tun-curr}), respectively.
Figures \ref{BEC-curr} shows
standard  and fast-forward (with a non-uniform time-scaling factor
with its mean $\bar{\alpha}=5$) cases, respectively.
Standard  tunneling current has a peak at $t=t_{0}\sim \frac{x_0}{v}$, when the soliton almost reaches the barrier.
The fast-forward tunneling current  is a
squeezed and enhanced version of the standard one. Since the soliton  arrives at the barrier earlier than standard arriving time, the peak of the current is realized at time $t_{0FF}= \frac{x_0}{\bar{\alpha}v}$.  Figure \ref{BEC-curr} also shows the enhancement of tunneling rate by
 $\bar{\alpha}=5$ as indicated by Eq.(\ref{ratio}).
In Fig.\ref{sol-drive} the driving electric field $E_{FF}$
necessary for the fast-forwarding of the soliton is
evaluated by Eq.(\ref{ele-fld-1d})  and is depicted as a function of $x$ and $t$.

\section{Conclusion}\label{sec-conclusion}
We developed a theory of fast-forwarding of quantum dynamics for charged particles, which exactly accelerates both amplitude and phase of the wave function throughout the fas-forward time range.
We elucidated the nature of the driving electro-magnetic field together with vector and scalar potentials to guarantee these exact fast forwarding.
The theory is applied to the tunneling phenomena through a tunneling
barrier. Typical examples described here are: 1) the initially-exponential wave packet moving
through the delta-function barrier; 2) the opened
Moshinsky shutter with a delta-function barrier just behind the
shutter. Standard (non-accelerated)
dynamics in these examples is known to be exactly solvable.
We see the remarkable squeezing and enhancement
of the tunneling current density, caused by
the fast-forwarding of quantum tunneling. We find: even if the barrier height will be increased,
one can generate a recognizable tunneling current by using a large enough time-scaling factor
$\alpha(t)$. At the same time, we have shown:
so long as $\alpha(t)$ is less than $\alpha_{max} \equiv \frac{V_0}{\mathcal{E}_0} (>1)$ with the barrier height $V_0$ and incident energy $\mathcal{E}_0$ in the standard tunneling,
the corresponding fast-forwarded dynamics is also the tunneling phenomenon
keeping the particle's energy $\mathcal{E}_{FF}(t)$ below the barrier height $V_0$ throughout the time evolution,
and the time scaling works more effectively for the particle with lower incident energy.
The analysis is also carried
out on the acceleration of macroscopic quantum tunneling with use
of the nonlinear Schr\"odinger equation which accommodates a
delta-function barrier.

Finally we should note that this work is inside a broader
concept to enhance the visibility of quantum transient phenomena (post-exponential decay\cite{tor,cam}, quantum backflow\cite{pal}, diffraction in
time\cite{ris,cor}, as well as quantum tunneling) which are predictable by
quantum mechanics but hardly detectable because  the detection number of
particles is very small. The general theory in Section \ref{new appro} will be an alternative vehicle to optimize the visibility parameters to improve those feeble observations.
\\{\em Acknowledgments.} One of the authors (K. N.) is grateful
to S. Masuda, A. del Campo, M. Nakayama and Y. Musakhanov for
enlightening discussions and comments in various stages of this work.

\end{document}